\documentclass[12pt]{article}
\usepackage{epsf}

\begin{document}

\title{An analysis of celestial pole offset observations in the
       free core nutation frequency band}
\author{Zinovy Malkin, Natalia Miller \\
   Pulkovo Observatory, St. Petersburg, Russia \\
   e-mail: malkin@gao.spb.ru}
\date{April 24, 2007}
\maketitle
\footnotetext{Presented at the EGU 4th General Assembly, Vienna, Austria, 15-20 April 2007}

\begin{abstract}
In this study, three empirical Free Core Nutation (FCN) models developed
to the present time, MHB2000, Malkin's and Lambert's ones, are compared on
the basis of representation of variations of the FCN amplitude and phase
predicted by these models.
It is possible to conclude, that the model of the author provides
the most realistic representation of the FCN variations.
However, the specified models are based on representation about single
FCN rotational mode. At the same time, some results of processing
of the VLBI observations made during last years, specify possible presence
of two close FCN periods. A theoretical explanation to presence of a second
FCN frequency FCN has been given by G.~Krasinsky in his theory of rotation
of the Earth with two-layer liquid core, ERA2005.
In the present work, for more detailed studying this phenomenon, the IVS time
series of celestial pole offset, and also those predicted by the ERA2005 theory,
have been investigated by means of several statistical methods which confidently
show presence of two fluctuations in nutational movement of an Earth's rotation
axis with the periods about $-452$ and $-410$ solar days.
\end{abstract}
\bigskip

\section{Known empirical FCN models}

\begin{figure}[ht]
\centering
\hbox{
\epsfclipon \epsfxsize=0.48\textwidth \epsffile{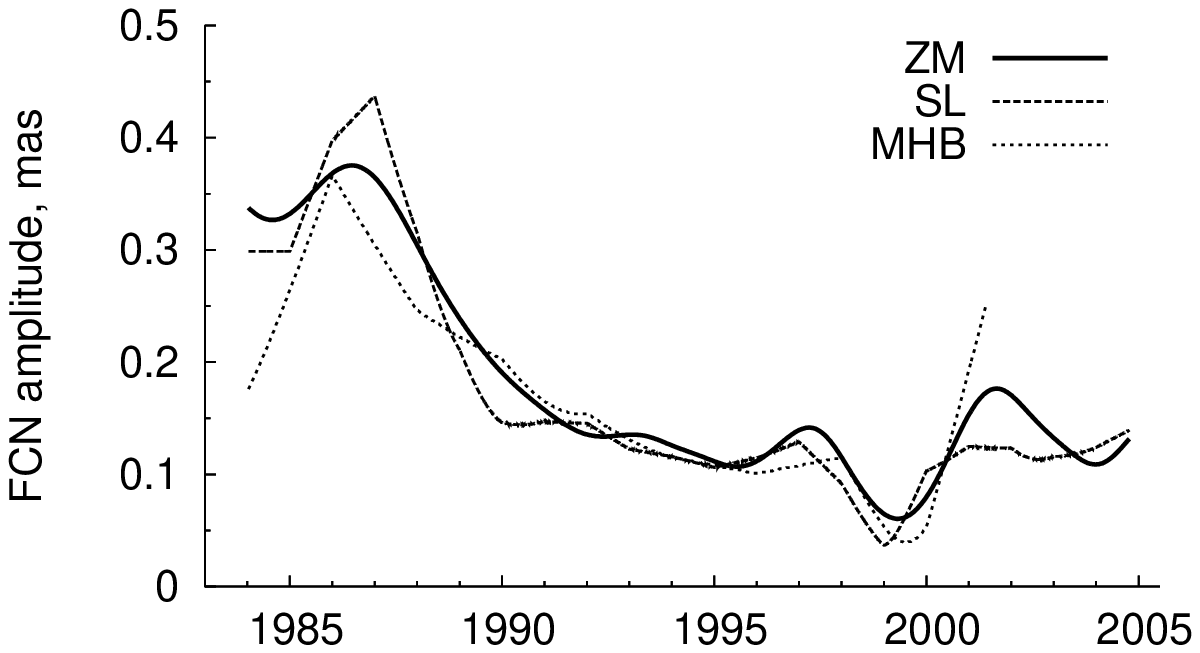}
\hskip 0.04\textwidth
\epsfclipon \epsfxsize=0.48\textwidth \epsffile{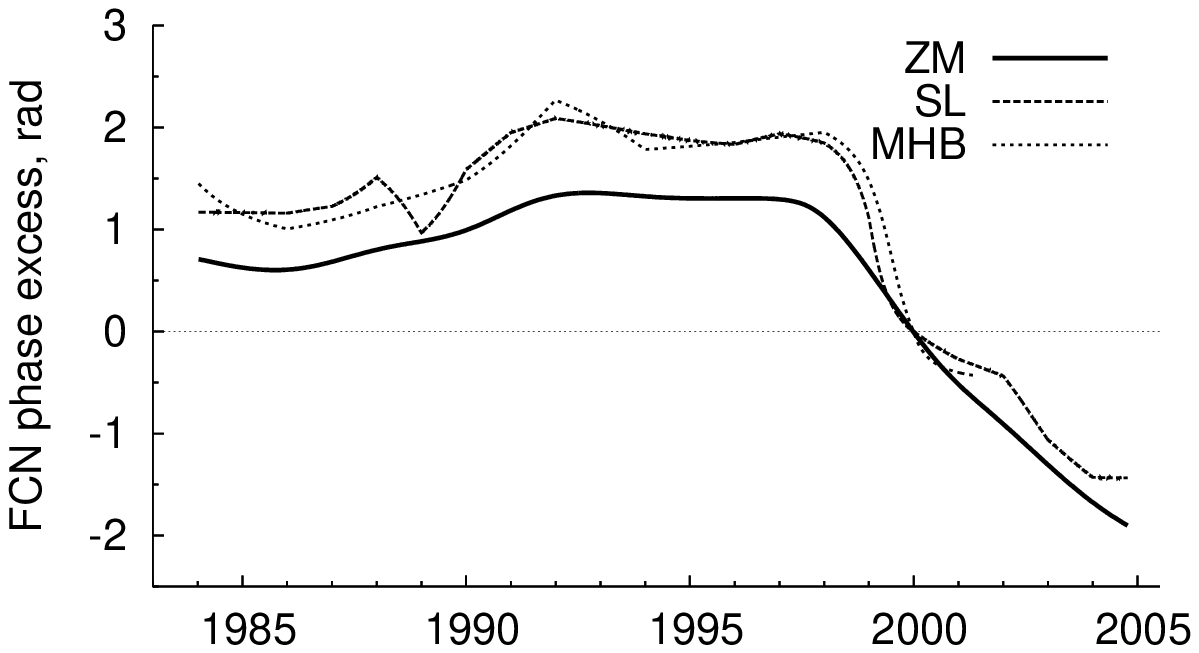}
}
\caption{\small Variation of the FCN amplitude and phase in accordance
  with three FCN models.}
\label{fig:fcnpar}
\end{figure}

To date,
three empirical FCN models have been developed: MHB2000 \cite{Herring02},
Malkin (ZM) \cite{Malkin03h}, Lambert (SL) \cite{McCarthy05}.
Comparison of these models with results of VLBI observations shows that
all three models
allow one to account equally essentially for the FCN contribution.
However, as can be seen from fig.~\ref{fig:fcnpar},
ZM model provides smooth and, apparently, more realistic representation
of variations of the basic geophysical parameters, the FCN amplitude
and phase.

It should be noted, however, that all models only approximate the observed
variations in the celestial pole offset, without striking into the FCN
physical properties.
Further development of FCN model can be reached by consideration of
a two-frequency FCN theory. Earlier, Malkin and Terentev
revealed a second fluctuation with the period about 410 solar days
in the celestial pole offset series \cite{Malkin03a, Malkin03c}.
However, then this fact has not been given due value.
Later, Schmidt {\it et al.} \cite{Schmidt05} have shown
the presence of two periods $-435$ and $-410$ days by means of the wavelet
analysis with high frequency resolution.
They supposed that observed variations in the FCN amplitude and phase
are caused by beating between two oscillations with close periods.
A theoretical explanation of this phenomenon has been given by Krasinsky
\cite{Krasinsky06a} in his numerical theory of rotation
of the Earth, ERA2005, considering two-layer structure of the fluid core.

In this paper, we performed more detailed analysis of the celestial pole offset
time series provided by the International VLBI Service for Geodesy and Astrometry
(IVS) for the period 1989--2006. Earlier observations were not analyzed
in view of their relatively low accuracy \cite{Malkin03h,Malkin04a}.
The smoothed differences between the observed celestial pole offset values
and those predicted by the IAU200A model were used in our study.

\section{Spectrum analysis}

\begin{figure}
\centering
\hbox{
\epsfclipon \epsfxsize=0.48\textwidth \epsffile{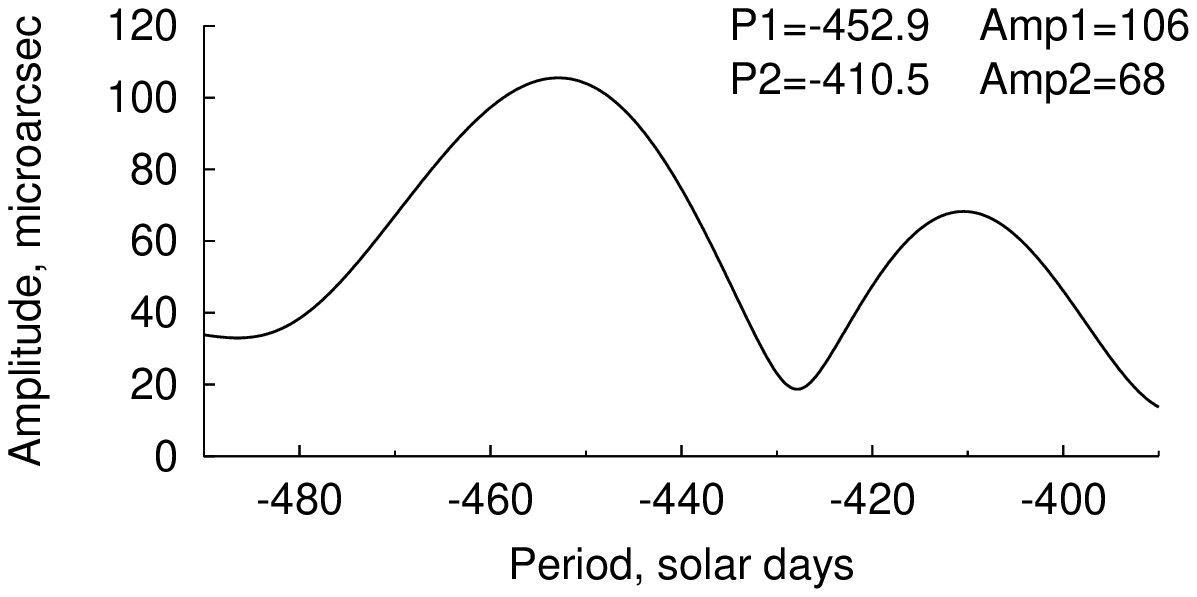}
\hskip 0.04\textwidth
\epsfclipon \epsfxsize=0.48\textwidth \epsffile{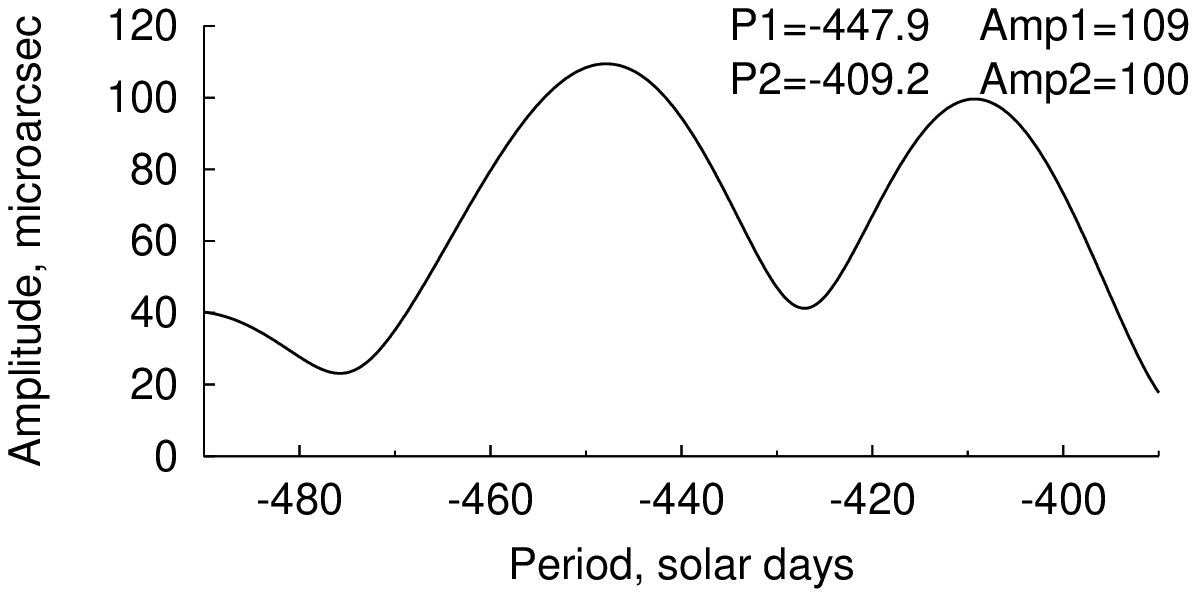}
}
\caption{\small Spectra of the IVS (left)and ERA2005 (right) celestial pole
  offset series.}
\label{fig:spectra}
\end{figure}

Firstly, we applied the discrete Fourier spectral analysis to the analyzed
time series. We computed the Schuster periodogram for complex series $X+iY$.
Usually for the spectrum analysis, the Fast Fourier Transform technique,
which provides calculation of the spectrum estimations on a grid
of frequencies, multiple to Nyquist frequency that does not provide the
detailed frequency resolution, is used. To increase the frequency
resolution, we used direct calculation of spectrum estimates, which
allowed us to use any, as much as dense grid of frequencies (periods).
Also, for increase of the frequency resolution, a frequency window
was not applied, which does not lead to deterioration of results
in this case, as initial data are smooth enough, and we study narrow
enough band of a spectrum.
The spectra of the differences between the IVS
celestial pole offset time series and those predicted by the
ERA2005 model are presented in Fig.~\ref{fig:spectra}, which shows
reasonably good agreement of the ERA2005 theory and observations.
Also in the investigated time series, the annual component is surely
revealed, but with essentially smaller amplitude, than the FCN components.

\section{Principal components analysis}

\begin{figure}
\centering
\epsfclipon \epsfxsize=0.9\textwidth \epsffile{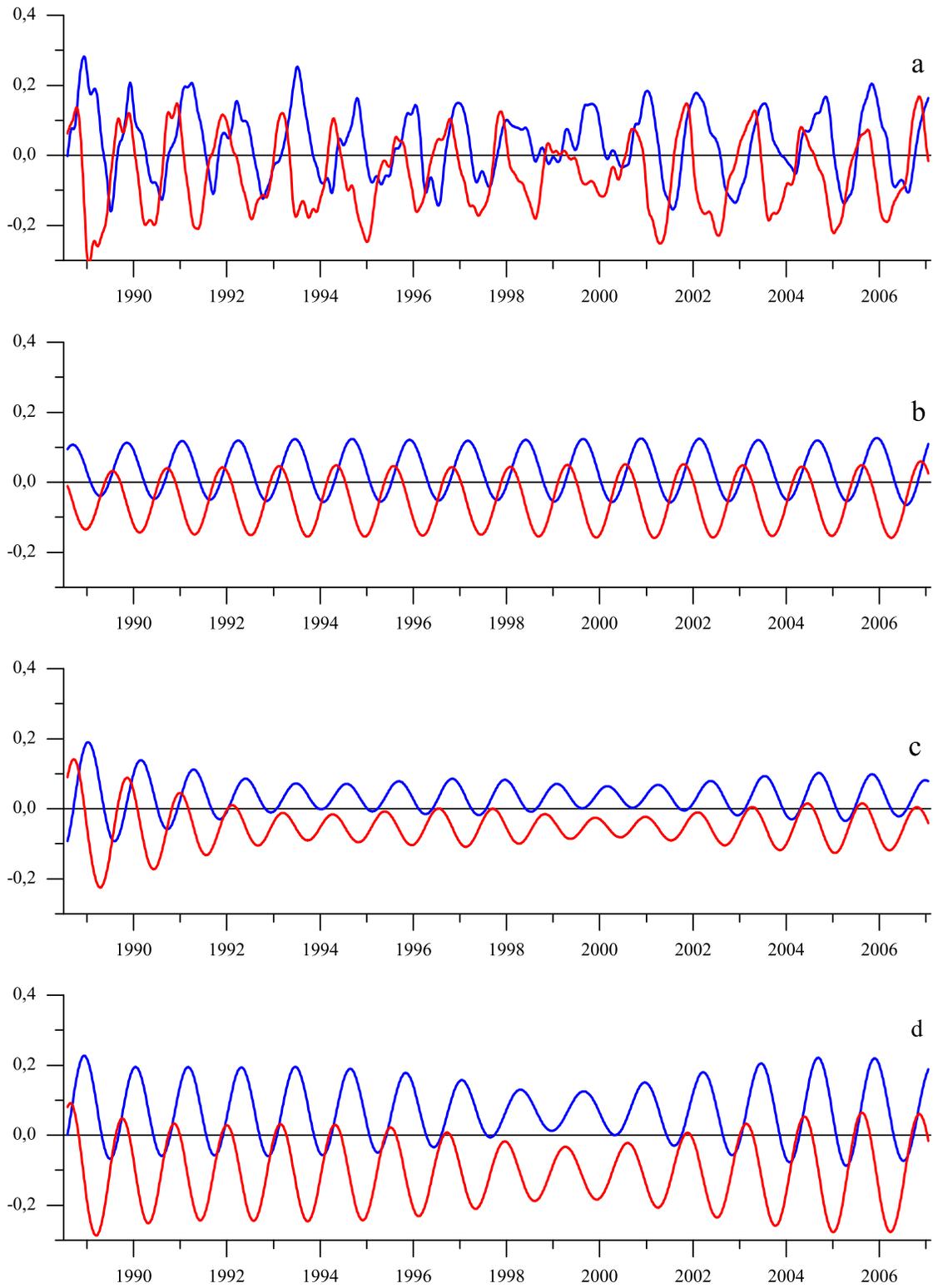}
\caption{\small Input time series and main harmonic principal components
obtained from the PCA (blue~-- X, red~-- Y): a~-- input, b~-- first principal
component with period 452 days,  c~-- second principal component with period
409 days, d~-- sum of two principal components.}
\label{fig:pc}
\end{figure}

The second method we used for investigation of the celestial pole offset series
was Principal Component Analysis (PCA) also known as Singular Spectrum Analysis.
The ``Caterpillar'' software developed at the St.~Petersburg State University
was used for computation ({\tt http://www.gistatgroup.com/cat/}).
It should be mentioned that, except harmonic components, PCA allow one
to isolate an actual long-term trend component not burdened by any
assumption about its {\it a priori} model.
Usually, the trend contribution in an analyzed time series as determined
from the PCA is large enough (40.5\% in our case), and it can distort the harmonic
components under investigation.
To mitigate this effect, we performed the computations in two iterations.
At the first iteration, all principal components were resolved, and
at second one, the main trend components found at the first iteration
were removed from the input time series.
Fig.~\ref{fig:pc} shows the original time series and two principal components
PC1 and PC2 found from the analysis.
Three most valuable harmonic components are those with periods 452 solar days
(the contribution is equal to 53.8\%), 409 days (19.0\%) and 366 days (6.8\%).
Specified period values were found from the spectral analysis made in the same way
as described in the previous section.

\section{Wavelet analysis}

\begin{figure}
\centering
\begin{tabular}{lc}
{\small IVS--IAU2000} &
   \begin{tabular}{c}
   \epsfclipon \epsfxsize=0.45\textwidth \epsffile{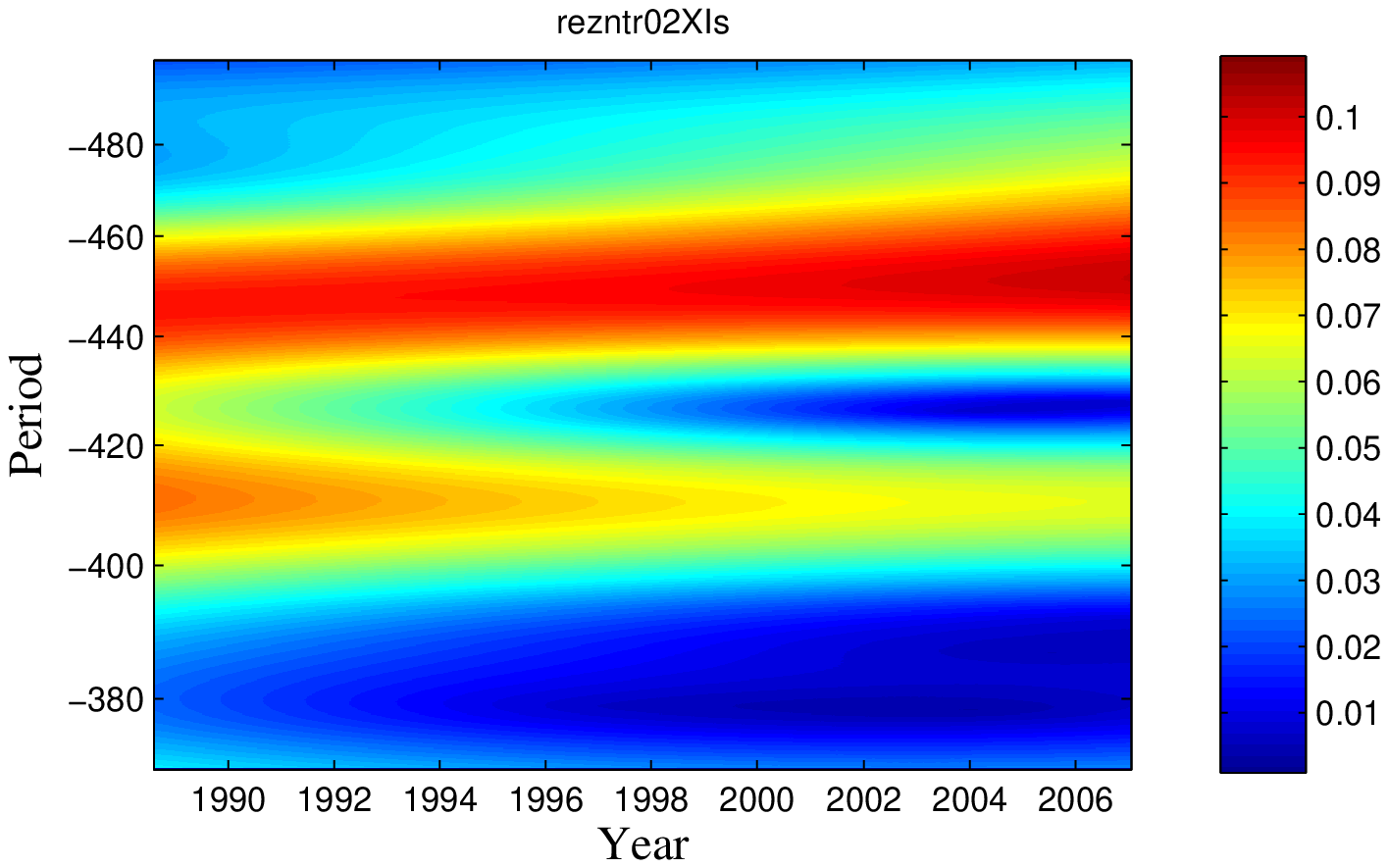} \\
   \end{tabular} \\
{\small PC1} &
   \begin{tabular}{c}
   \epsfclipon \epsfxsize=0.45\textwidth \epsffile{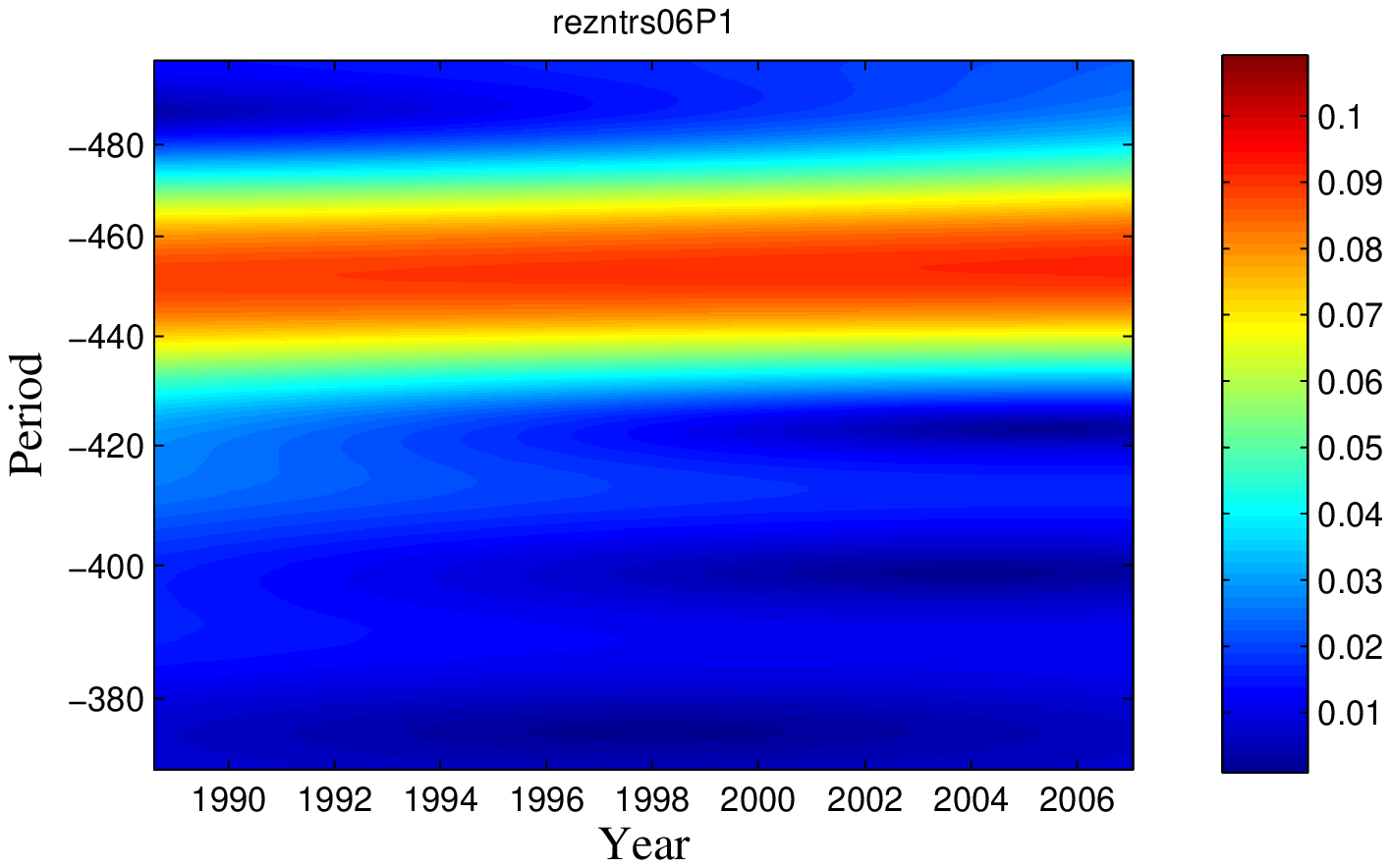} \\
   \end{tabular} \\
{\small PC2} &
   \begin{tabular}{c}
   \epsfclipon \epsfxsize=0.45\textwidth \epsffile{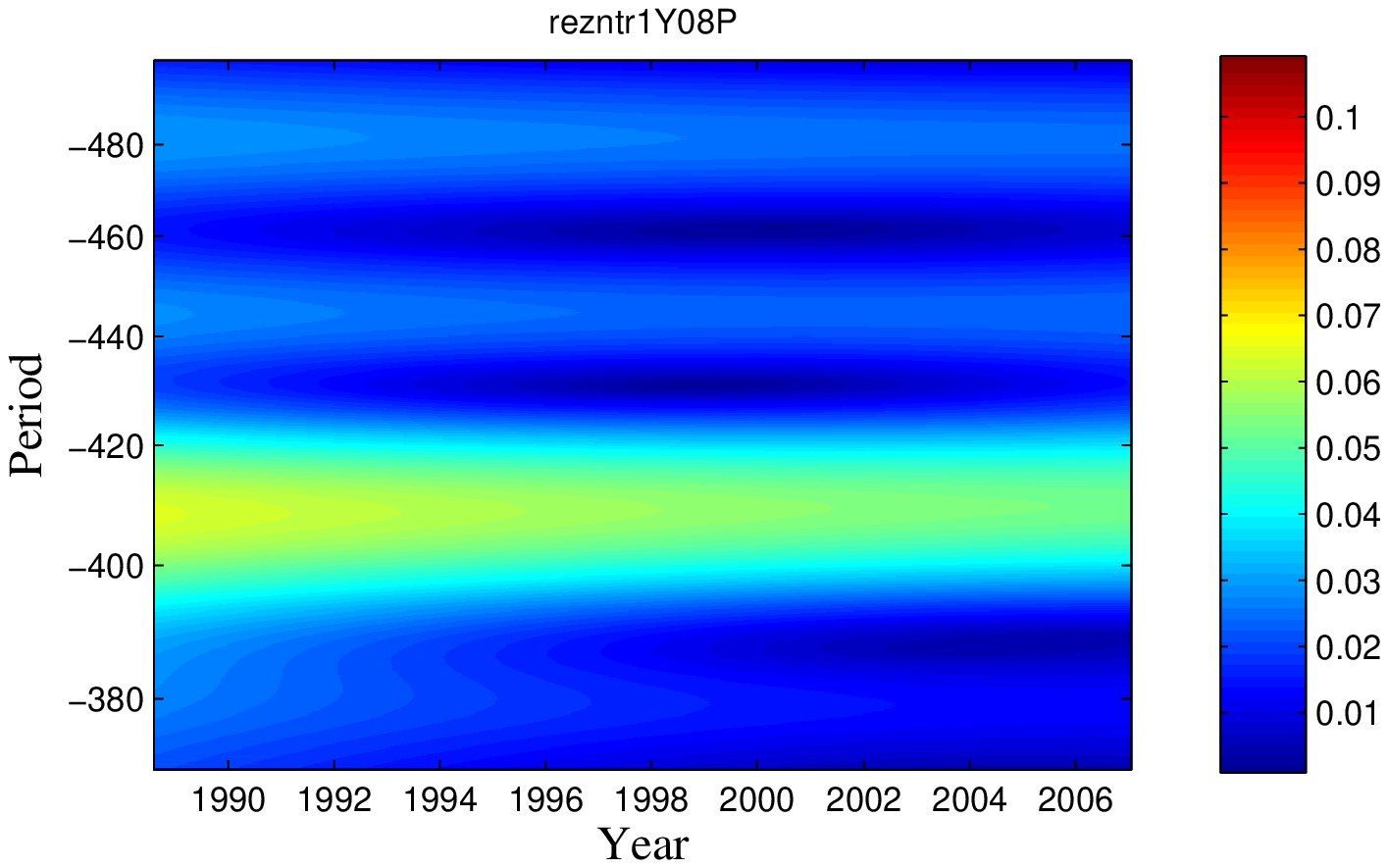} \\
   \end{tabular} \\
{\small PC1 + PC2} &
   \begin{tabular}{c}
   \epsfclipon \epsfxsize=0.45\textwidth \epsffile{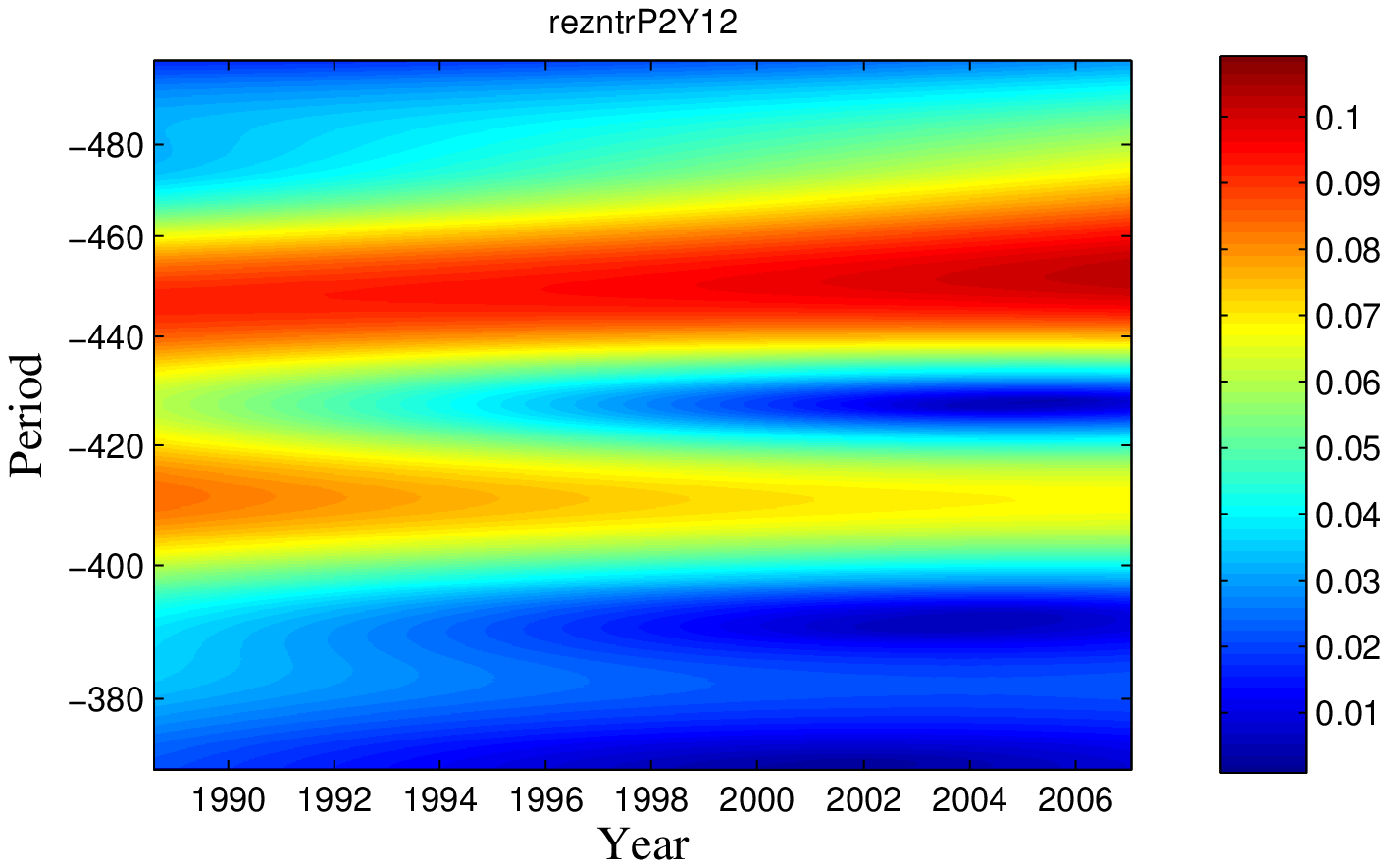} \\
   \end{tabular} \\
{\small ERA2005} &
   \begin{tabular}{c}
   \epsfclipon \epsfxsize=0.45\textwidth \epsffile{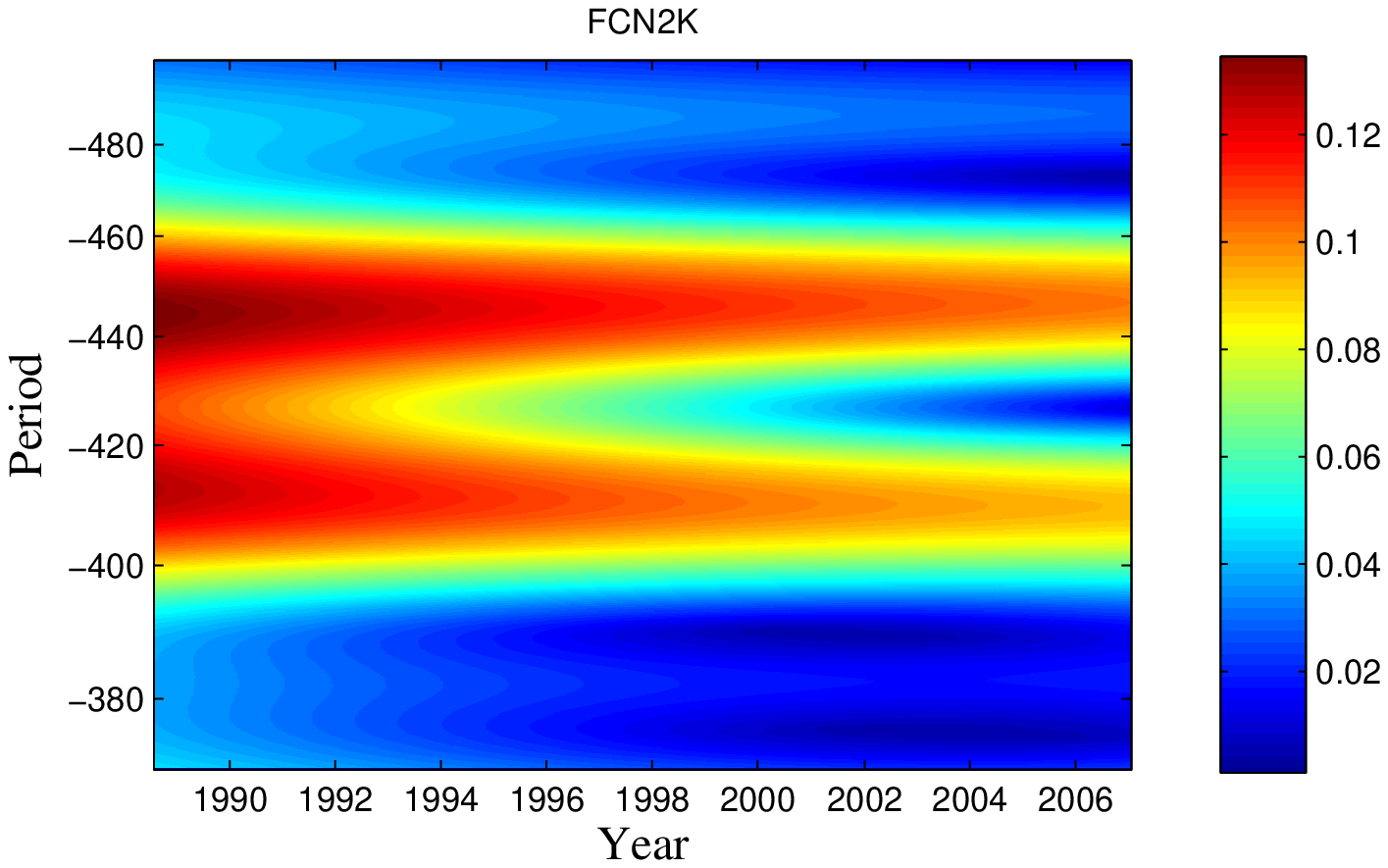} \\
   \end{tabular} \\
\end{tabular}
\caption{\small Results of the wavelet analysis of the input time series,
its principal components and ERA2005 series. Amplitude scale is given in mas.}
\label{fig:wavelet-fcn}
\end{figure}

To investigate how the two FCN components vary with time,
the wavelet technique was applied to the IVS celestial pole offset,
two harmonic principal components found from the PCA and
ERA2005 time series.
The WWZ software ({\tt http://www.aavso.org/})
developed at the American Association of Variable Star Observers
was used for analysis.
The mathematical background of this method is described in \cite{Foster96c}.
Results of the wavelet analysis are presented in Fig.~\ref{fig:wavelet-fcn}.
Two periodic components can be clearly seen at both observed and
theoretical time series.
For better comparison,
we applied the wavelet with the same parameter $\sigma$=10
as was used in \cite{Schmidt05}, which provides high frequency resolution.
The time resolution is rather pure in such a case however.
So, wavelet estimates,in fact, are averaged for many-year interval.

\section{Discussion and conclusion}

In this paper, we investigated the IVS celestial pole offset time series
as well as theoretical ERA 2005 time series by means of
three statistical tools, Discrete Fourier Transform, Principal
Component Analysis and wavelet analysis, in the FCN frequency band.
The results obtained with all the methods definitely show presence
of two harmonic components with periods about $-410$ and $-452$ solar days,
yet these methods are not fully independent.

This result confirms ones obtained in \cite{Malkin03a,Malkin03c,Schmidt05}
from VLBI data processing and the ERA2005 theory
\cite{Krasinsky06a,Krasinsky06b}.
However, the values of the FCN component periods found here and
obtained in the previous papers are substantially different.
Moreover, supplement study has shown that the period of FCN components
depend on the time span of data used for analysis.
So, further investigations are needed before making a final conclusion.
Also, it's important to compare these results with the resonance FCN
period, see e.g. \cite{Vondrak05}, in particular, considering
a two-component resonance model.

Finally, we can hope that using two-component empirical FCN model,
in case it is proved to be real, will allow us to predict the FCN
contribution to the nutation series with better accuracy than
existing models.

\end{document}